\pdfoutput=1

\documentclass{sig-alternate-2013-tweeked}

\usepackage{nicefrac}
\usepackage{listings}
\usepackage{graphicx}
\usepackage[tight]{subfigure}
\usepackage[usenames]{color}
\usepackage{float}
\usepackage{multirow}
\usepackage{enumitem}
\usepackage{ifpdf}
\usepackage{hyperref}
\usepackage{breakurl}

\mathchardef\UrlBreakPenalty=10

\newcommand{\squishlist}{
  \begin{list}{$\bullet$}
  {
   \setlength{\itemsep}{0pt}
   \setlength{\parsep}{0pt}
   \setlength{\topsep}{0pt}
   \setlength{\partopsep}{0pt}
   \setlength{\leftmargin}{2em}
   \setlength{\labelwidth}{1.5em}
   \setlength{\labelsep}{0.5em} } }

\newcommand{\squishlisttwo}{
  \begin{list}{$\bullet$}
  {
   \setlength{\itemsep}{0pt}
   \setlength{\parsep}{6pt}
   \setlength{\topsep}{6pt}
   \setlength{\partopsep}{0pt}
   \setlength{\leftmargin}{2em}
   \setlength{\labelwidth}{1.5em}
   \setlength{\labelsep}{0.5em} } }

\newcommand{\squishend}{
   \end{list}  }

\newfont{\mycrnotice}{ptmr8t at 7pt}
\newfont{\myconfname}{ptmri8t at 7pt}
\let\crnotice\mycrnotice%
\let\confname\myconfname%

\permission{}
\conferenceinfo{ }{{ }}
\copyrightetc{Copyright 2013 Scott G. Ainsworth and Michael L. Nelson}
\crdata{978-1-4503-2077-1/13/07} 
\boilerplate{}
\begin{document}


\title{Evaluating Sliding and Sticky Target Policies by Measuring
       Temporal Drift in Acyclic Walks Through a Web Archive}

\numberofauthors{2}
\author{
\alignauthor
Scott G. Ainsworth\\
       \affaddr{Old Dominion University}\\
       \affaddr{Norfolk, VA, USA}\\
       \email{sainswor@cs.odu.edu}
\alignauthor
Michael L. Nelson\\
       \affaddr{Old Dominion University}\\
       \affaddr{Norfolk, VA, USA}\\
       \email{mln@cs.odu.edu}
}

\maketitle

\begin{abstract}

When a user views an archived page using the archive's user
  interface (UI), the user selects a datetime to view from a list.
The archived web page, if available, is then displayed.
From this display, the web archive UI attempts to simulate the web browsing
  experience by smoothly transitioning between archived pages.
During this process, the target datetime changes with each link followed;
  drifting away from the datetime originally selected.
When browsing sparsely-archived pages, this nearly-silent drift can be
  many years in just a few clicks.
We conducted 200,000 acyclic walks of archived pages, following up to 50 links
  per walk, comparing the results of two target datetime policies.
The Sliding Target policy allows the target datetime to change as it does in
  archive UIs such as the Internet Archive's Wayback Machine.
The Sticky Target policy, represented by the Memento API, keeps the target
  datetime the same throughout the walk.
We found that the Sliding Target policy drift increases with the number of
  walk steps, number of domains visited, and choice (number of links
  available).
However, the Sticky Target policy controls temporal drift, holding it
  to less than 30 days on average regardless of
  walk length or number of domains visited.
The Sticky Target policy shows some increase as choice increases, but this
  may be caused by other factors.
We conclude that based on walk length, the Sticky Target policy generally
  produces at least 30 days less drift than the Sliding Target policy.
\end{abstract}




\section{Introduction}
\label{s:Introduction}

To browse archived pages from a web archive such as the Internet Archive
  \cite{carpenter-ndiipp10}, the user begins with the selection of a URI
  followed by selection of a Memento-Datetime (the datetime the resource was
  archived).
Following these selections, the user is able to browse the archive's collection
  of mementos (archived copies) by clicking links on displayed pages;
  a process similar to browsing the live Web.
However, with each click, the target datetime (the datetime requested by the
  user) is changed to the Memento-Datetime of the displayed page.
Although this constant change is visible in the web browser address bar and the
  archive's user interface (UI), the change is easy to overlook because the
  change happens without explicit user interaction.

\begin{figure*}
  \centering  
  \subfigure[UI CS 2005-05-14]{
    \includegraphics[width=2.21in]{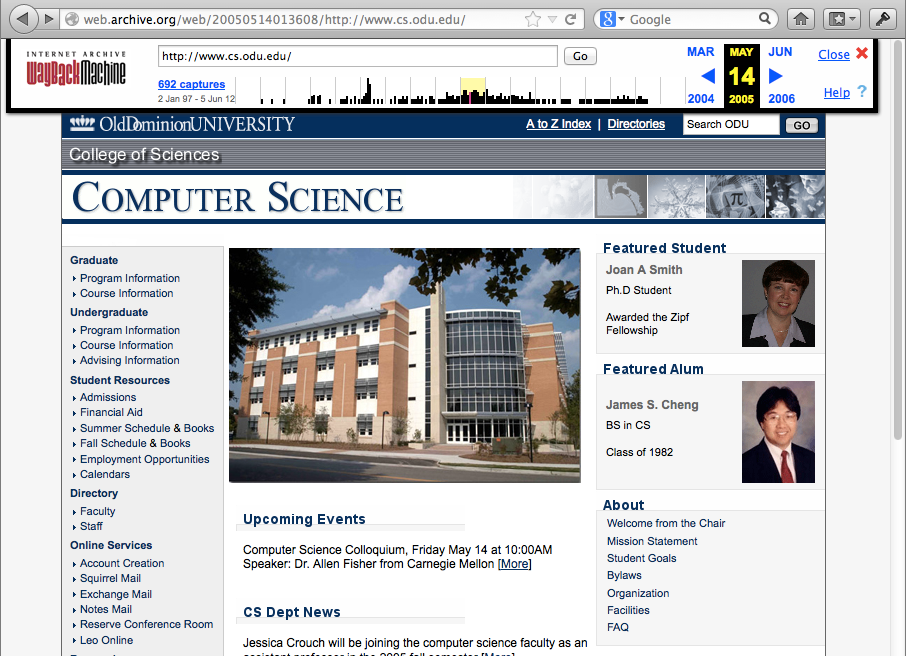}
    \label{f:oducsdrift:ui:cshome:a}
  }
  \subfigure[UI Sci 2005-04-22]{
    \includegraphics[width=2.21in]{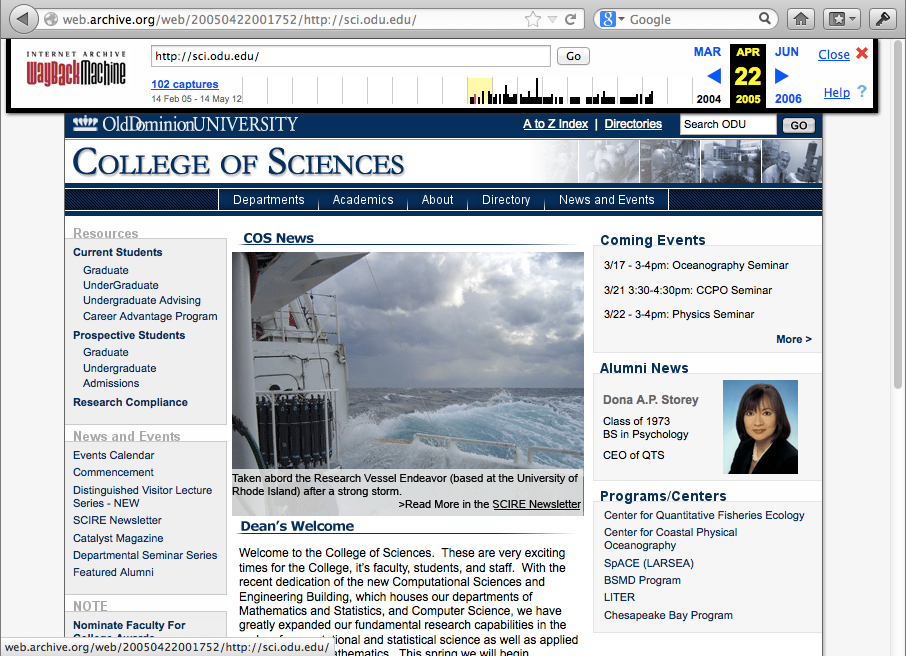}
    \label{f:oducsdrift:ui:csihome}
  }
  \subfigure[UI CS 2005-03-31]{
    \includegraphics[width=2.21in]{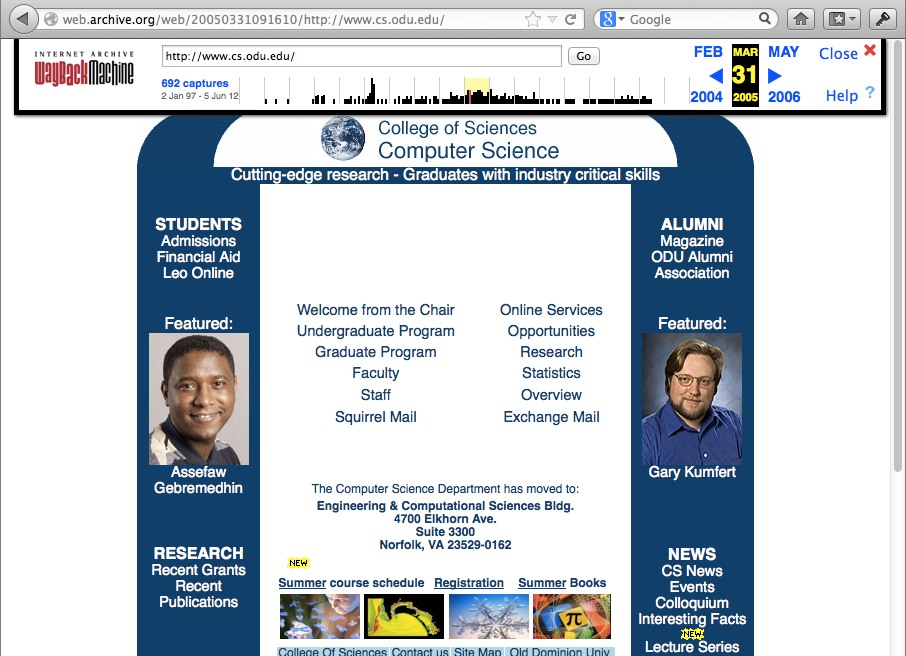}
    \label{f:oducsdrift:ui:cshome:b}
  }
  \subfigure[API CS 2005-05-14]{
    \includegraphics[width=2.21in]{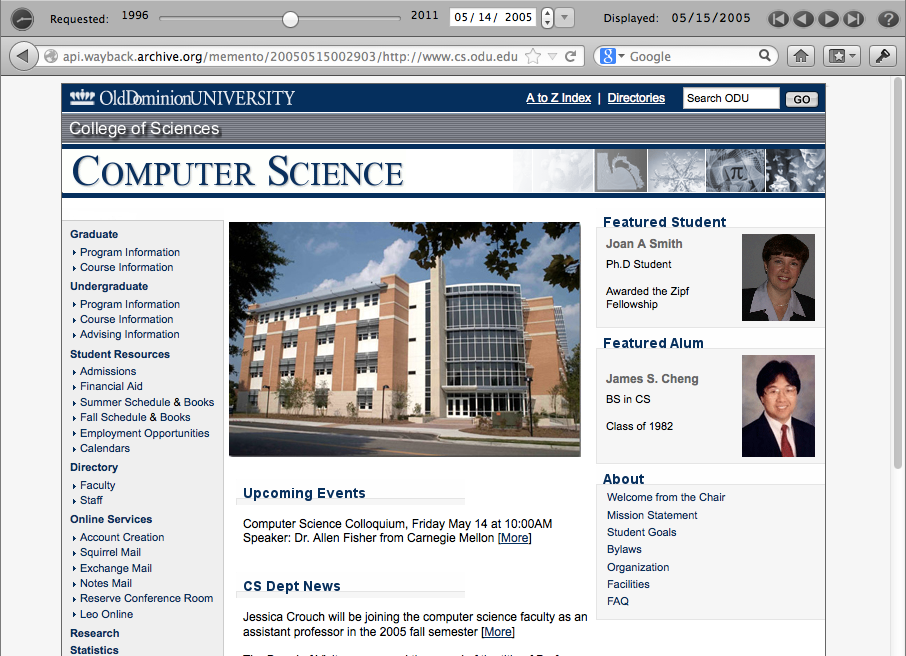}
    \label{f:oducsdrift:api:cshome:a}
  }
  \subfigure[API Sci 2005-04-22]{
    \includegraphics[width=2.21in]{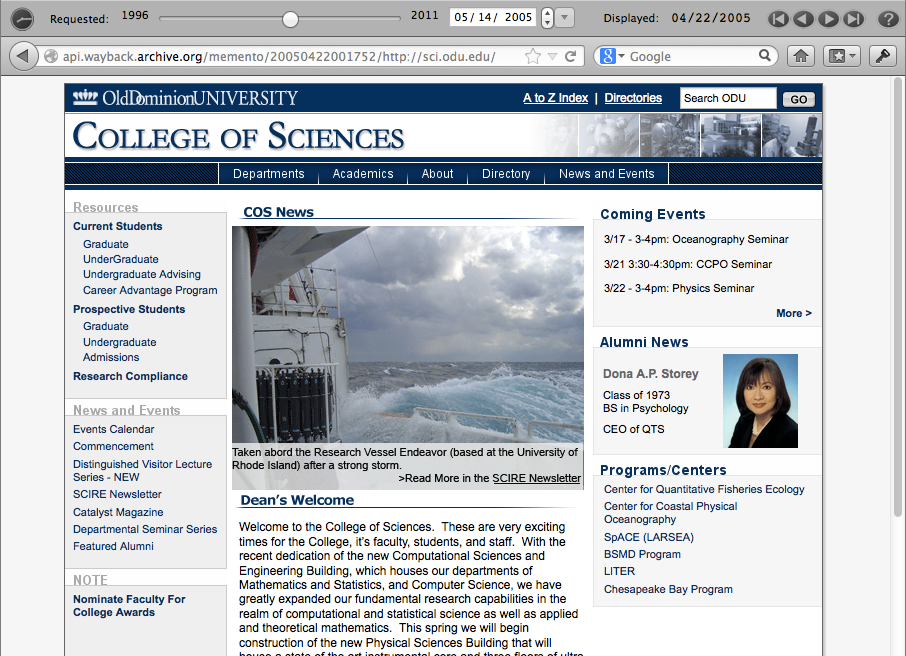}
    \label{f:oducsdrift:api:csihome}
  }
  \subfigure[API CS 2005-05-14]{
    \includegraphics[width=2.21in]{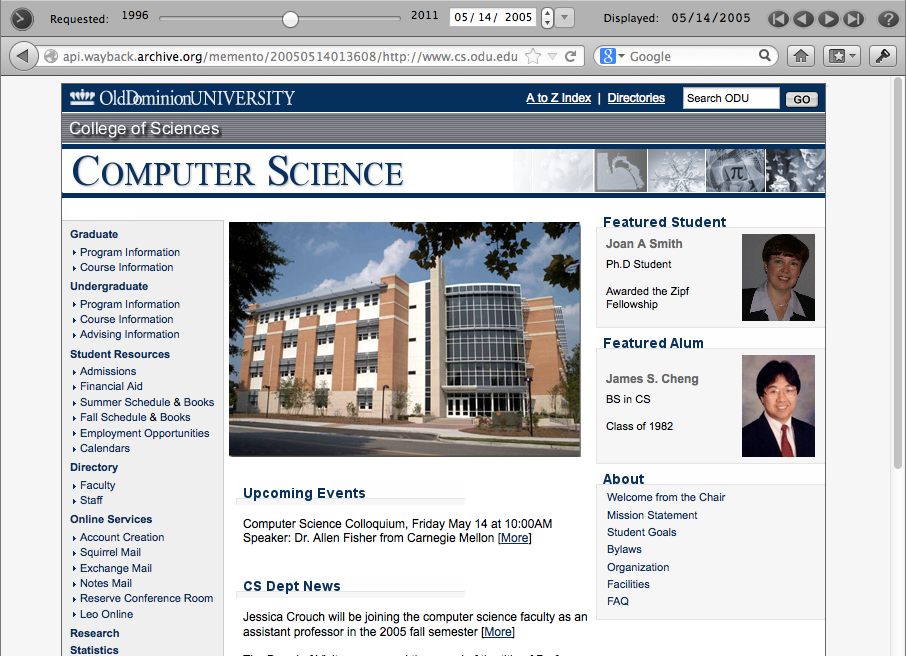}
    \label{f:oducsdrift:api:cshome:b}
  }
  \caption{Impact of Drift on Archive Browsing}
  \label{f:oducsdrift}
\end{figure*}

The screen shots in the top row of Figure \ref{f:oducsdrift}
  illustrates a clear case of this phenomenon.
Archives of the Old Dominion University Computer Science and College of Sciences
  home pages are shown.
The process begins by entering \url{http://www.cs.odu.edu} in the Internet
  Archive's archive browser, The Wayback Machine.
The user is then presented with a list of archive datetimes,
  from which May 14, 2005 01:36:08 GMT is selected.
The user views the Computer Science home page
  [Figure \ref{f:oducsdrift:ui:cshome:a}].
The page URI is
  \url{http://web.archive.org/web/20050514013608/http://www.cs.odu.edu/};
  note that the datetime encoded\footnote{Date and time formatted YYYYMMDDHHMMSS}
  in the URI matches the date selected.
When the user clicks the {\it College of Sciences} link, the page is displayed
  [Figure \ref{f:oducsdrift:ui:csihome}].
However, the encoded datetime changed to 20050422001752, a drift of 22 days.
  This datetime also becomes the new target datetime.
When the user clicks the {\it Computer Science} link, the result is a different
  version than first displayed, as shown in
  Figure \ref{f:oducsdrift:ui:cshome:b}.

On the other hand, using a Memento-enabled browser, such as Firefox
  with the MementoFox add-on \cite{sanderson-code4lib11}, maintains
  a consistent target datetime as the user follows links.
The bottom row of Figure \ref{f:oducsdrift} shows the results.
Using the API,
  each visit to the Computer Science home page returns the same version
  [Figures \ref{f:oducsdrift:api:cshome:a} and \ref{f:oducsdrift:api:cshome:b}].
The rough statistics in Table \ref{t:ExampleDrift} show the potential
  improvement that can be achieved using the Memento API.

\begin{table}[h]
\centering
\caption{Temporal Drift Example}
\label{t:ExampleDrift}
\begin{tabular}{l|c@{ }r@{}|c@{ }r}
  \hline
    & \multicolumn{2}{c}{Wayback Machine UI}
    & \multicolumn{2}{|c}{Memento API} \\
  \cline{2-5}
  Page & Datetime & \multicolumn{1}{c|}{Drift}
       & Datetime & \multicolumn{1}{c}{Drift} \\
  \hline
  CS Home  & 2005-05-14 & \multicolumn{1}{c|}{--} & 2005-05-14 & \multicolumn{1}{c}{--} \\
  Sci Home & 2005-04-22 & 22 days~~ & 2005-04-22 & ~~22 days \\
  CS Home  & 2005-03-31 & 44 days~~ & 2005-05-14 & 0 days \\
  \hline
  Mean     &  & 33 days~~ &  & 11 days \\
  \hline
\end{tabular}
\end{table}

The simple example above raises many questions.
How much drift do users experience when browsing archives using user interfaces
  such as the Wayback Machine?
If the Memento API is used instead, how much drift is experienced?
Which method is better and by how much?
What factors contribute, positively or negatively, to the amount of drift?
  In particular, does the number of links available (choice), number of domains
  visited, or the number of links followed (walk length) contribute to drift?

\section{Related Work}
\label{s:Related_Work}


Although the need for web archiving has been understood since nearly the dawn
  of the Web \cite{casey-crl98}, these efforts have been for the most part
  independent in motivation, requirements, and scope.
The Internet Archive, the first archive to attempt global scope, came into
  existence in 1995 \cite{kimpton-webarchiving06}.
Since then, many other archives have come into existence.
Some of these use software developed by the Internet Archive and have similar
  capture behavior and user interfaces;
however, other archives such as WebCite \cite{eysenbach-jmir05} have
  significantly different capture behaviors.

Large-scale web archiving requires resolution of issues and approaches on
  several axes.
Although somewhat out of date, Masan\`es \cite{masanes-webarchiving06} is an
  excellent introduction.
Masan\`es covers a broad range of web archiving topics.
Of significance to this research are the technical aspects of acquisition,
  organization and storage, and quality and completeness.
A major area not addressed by Masan\`es is access to archives, in particular
  the lack of standards or conventions for accessing archived resources.
Van de Sompel et al.\ \cite{vandesompel-arxiv:0911.1112} addressed this lack with
  Memento.

\subsection{Acquisition}
\label{ss:Acquisition}

Acquisition is the technical means of bringing content into an archive.
  Client-side archiving essentially emulates web users following links,
  obtaining content using the HTTP protocol.
The Heritrix \cite{mohr-iwaw04} crawler and the mirroring capability of
  \textit{wget}\footnote{\url{http://www.gnu.org/software/wget/}} are examples
  of client-side archiving.
A significant issue with client-side archiving is that only those
  parts of the Web exposed as linked resources are captured.
  Transactional archiving is specifically designed to overcome this limitation.
Transactional archiving
  \cite{brunelle-arXiv:1209.1811,dyreson-www04,fitch-awww03}
  inserts the capture process between the user and the
  data source, for example an Apache web server filter, which requires the
  cooperation of the server operator.
Unique request-response pairs are archived, including requests for resources
  that are not linked.
Server-side archiving makes a direct copy of the content from the server,
  bypassing HTTP altogether.
Although conceptually simple, access to the resulting
  server-side archive can be difficult,
  requiring different URIs and navigational structures than the original.
Many systems, e.g. content management systems and wikis, perform server-side
  archiving by design.
  
\subsection{Organization and Storage}
\label{ss:OrganizationAndStorage}

Once acquired content must be stored.
Masan\`es \cite{masanes-webarchiving06} describes three organization and storage
  methods that are commonly used. 
  Local archives store content in the local file system, transforming the
  content just enough to allow off-line browsing.
Links must be modified to reference either locally-stored archived resources
  or the live web.
Strict adherence to the original content is generally impractical and size is
  limited by local storage capacity and speed.
Thus, local archives are most suitable~for small-scale archiving.
A common method of creating local archives is \textit{wget} mirroring.
  Web-served archives, like the IA,
  commonly store content in WARC (Web ARChive) container
  files, which allows the original content and URIs to be stored unmodified.
This also overcomes many limitations imposed by file systems.
Content is provided to users over HTTP.
Web-served archiving is highly scalable and suitable for large-scale archiving.
Non-web archives generally transform web content into other forms.
For example, Adobe Acrobat has the ability to download web content and produce
  a corresponding PDF.
This type of archiving is generally best suited for resources, such as digitized
  books, originally created independently from the Web.
Of the three types of organization and storage methods, only web-served archives are
  relevant to this study.

\subsection{Access}
\label{ss:Access}

An area of web archives that remained unresolved until recently was lack of
  methods or a standard API for time-based access to archived resources.
Each archive provides a user interface (UI) to access the archive's resources.
(Many archive's use the Internet Archive's Wayback Machine
  \cite{tofel-iwaw07} and therefore share similar UIs.)
In general, UI access to archives starts with a user-selected URI
  and datetime, after which the archive allows the user
  to simply click links to browse the collection.
UI archive access is addressed in greater detail in Section
\ref{ss:BrowsingViaTheWaybackMachine}.

Van de Sompel et al.\ addressed the lack of a standard API with Memento
  \cite{draft-vandesompel-memento, vandesompel-arxiv:0911.1112}, an HTTP-based
  framework that bridges web archives with current resources.
It provides a standard API for identifying and dereferencing archived resources
  through datetime negotiation.
In Memento, each original resource, $\text{URI-R}$,  has zero or more archived
  representations, $\text{URI-M}_{i}$, that encapsulate the URI-R's state at
  times $t_{i}$.
Using the Memento API, clients are able to request $\text{URI-M}_{i}$ for a
  specified URI-R by datetime.
Memento is now an IETF Internet Draft \cite{draft-vandesompel-memento}.
Memento archive access is addressed in greater detail in Section
  \ref{ss:BrowsingViaMemento}.

\subsection{Quality and Completeness}
\label{ss:QualityAndCompleteness}

In general, quality is functionally defined as fitting a particular use and
  objectively defined as meeting measurable characteristics.
This examination of web archive content is concerned with the latter.
For web archives, most quality issues stem from the difficulties inherent in
  obtaining content using HTTP \cite{masanes-webarchiving06}.
Content is not always available when crawled, leaving gaps in the coverage.
Web sites change faster than crawls can acquire their content, which leads to
  temporal incoherence.
Ben Saad et al.\ \cite{bensaad-tpdl11} note that quality and completeness
  require different methods and measures \textit{a priori} or \textit{a
  posterior}, that is during acquisition or during post-archival access
  respectively.

\subsubsection{Completeness (Coverage)}

When crawling to acquire content,
  the tradeoffs required and conditions encountered
  lead to incomplete content or coverage.
A web archive may not have the resources to acquire and store all content
  discovered.
Associated compromises include acquiring only high priority content and
  crawling content less often.
The content to be acquired may not be available at crawl time due to server
  downtime or network disruption.
The combination of compromises and resource unavailability create undesired,
  undocumented gaps in the archive.

Although much has been written on the technical, social, legal, and political
  issues of web archiving; little detailed research has been conducted on the
  archive coverage provided by the existing archives.
Day \cite{day-ecdl03} surveyed a large number of archives as part of
  investigating the methods and issues associated with archiving.
Day however does not address coverage.
Thelwall touches on coverage when he addresses international bias in the
  Internet Archive \cite{thelwall-lisr04}, but does not directly address how
  much of the Web is covered.
McCown and Nelson address coverage \cite{mccown-ist07}, but their research is
  limited to search engine caches. Ben
  Saad et al.\ \cite{bensaad-dexa11,bensaad-jcdl11} address qualitative
  completeness through change detection to identify and archive important
  changes (rather than simply archiving every change).
This research primarily addresses \textit{a priori} completeness.
\textit{A posteriori} web archive coverage is addressed by
  Ainsworth et al.\ \cite{ainsworth-jcdl11}.
Leveraging the Memento API and pilot infrastructure,
  Ainsworth et al.\ \cite{ainsworth-jcdl11} obtained results showing that
  35--90\% of publicly-accessible URIs have at least one publicly-accessible
  archived copy, 17--49\% have two to five copies, 1--8\% have six to ten
  copies, and 8--63\% at least ten copies. The number of URI copies varies as
  a function of time, but only 14.6--31.3\% of URIs are archived more than
  once per month.
The research also shows that coverage is dependent on social popularity.

\subsubsection{Temporal Coherence}

When crawling to acquire content, tradeoffs are required.
Crawling consumes server resources, thus crawls must be polite, e.g. paced to
  avoid adversely impacting the server.
The web archive may not have the bandwidth needed to crawl quickly.
These and other constraints increase crawl duration, which in turn increases
  the likelihood of temporal incoherence.

Spaniol et al.\ \cite{spaniol-wicow09} note that crawls may span hours or
  days, increasing the risk of temporal incoherence, especially for large
  sites, and introduces a model for identifying coherent sections of archives,
  which provides a measure of quality.
Spaniol et al.\ also present a crawling strategy which helps minimize incoherence
  in web site captures. 
In a separate paper, Spaniol et al.\ \cite{spaniol-iwaw09} also develop
  crawl and site coherence visualizations.
Spaniol's work, while presenting an \textit{a posteriori} measure, concerns
  the quality of entire crawls.

Denev et al.\ present the SHARC framework \cite{denev-vldb09}, which introduces
  a stochastic notion of \textit{sharpness}.
Sites changes are modeled as Poisson processes with page-specific change rates.
Change rates can differ by MIME type and depths within the site.
This model allows reasoning on the expected sharpness of an acquisition crawl.
From this they propose four algorithms for site crawling.
Denev's work focuses on \textit{a priori} quality of entire crawls and does
  not address the quality of existing archives and crawls.

Ben Saad et al.\ \cite{bensaad-tpdl11} address both \textit{a priori} and
  \textit{a posteriori} quality.
Like Denev et al.\ \cite{denev-vldb09}, the \textit{a priori} solution is
  designed to optimize the crawling process for archival quality.
The \textit{a posteriori} solution uses information collected by the
  \textit{a priori} solution to direct the user to the most coherent
  archived versions.

All of the above research shares a common thread: evaluation and control
  of completeness and temporal coherence during the crawl with the goal
  of improving the archiving process.
In contrast, our research takes a detailed look at the quality and use
  of existing archives.

\section{Browsing and Drift}
\label{s:Drift}

Fundamentally, drift is the difference between the target datetime
  originally required and the Memento-Datetime returned
  by an archive.
Drift can be forward or backward in time;
  in this study only the amount of drift is relevant.
This paper examines two target datetime policies:

\squishlist
  \item \textbf{Sliding Target}: the target datetime changes as the user
        browses.  The Memento-Datetime of the current page becomes the new
        target datetime.
  \item \textbf{Sticky Target}: the target datetime is set once at the beginning
        of the browsing session.
\squishend

\subsection{Sliding Target (The Wayback Machine)}
\label{ss:BrowsingViaTheWaybackMachine}

Browsing using the Internet Archive's Wayback Machine User Interface (UI)
  employs the \textit{Sliding Target} datetime policy.
This policy has the potential to introduce permanent drift at every step.
Here is the browsing process in detail:

\begin{enumerate}

\item \textbf{Select URI-R}.
Navigate to \url{http://www.archive.org} and enter a URI-R.
Clicking the \textit{Take Me Back} button displays a calendar of the most
  recent year for which the URI-R is archived.
The 2005 calendar for the ODU Computer Science home is shown in
  Figure \ref{f:wayback-calendar}.

\item \textbf{Select Memento-Datetime}.
Dates covered by blue circles have mementos for the URI-R.
Larger circles indicate multiple mementos for the same date.
Hovering over a circle pops up a box that allows mementos to be selected.
When a memento is selected, its Memento-Datetime becomes the target datetime
  and the corresponding memento it is displayed (as was previously shown
  in Figure \ref{f:oducsdrift}).
Drift is introduced when the selected memento redirects to another memento
  that has a different Memento-Datetime than originally selected.

\begin{figure}[h!]
  \centering
  \includegraphics[width=3.25in]{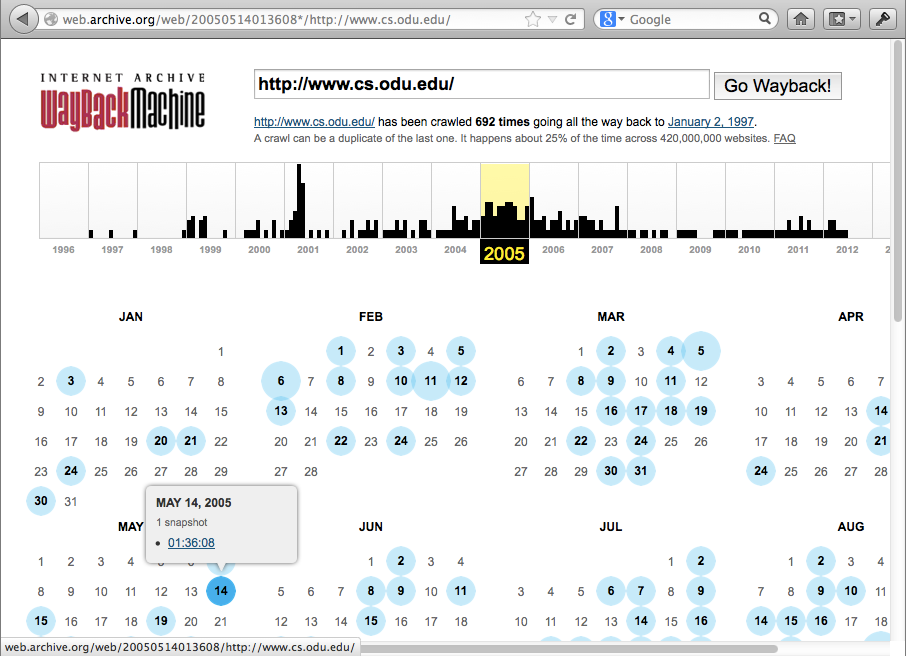}
  \caption{Wayback Machine Calendar}
  \label{f:wayback-calendar}
\end{figure}

\item \textbf{Browse additional URI-Rs}.
To simulate browsing the Web within the context of the archive, 
  links are rewritten to reference the archive instead of the
  original URI
  and to embedded the Memento-Datetime of the displayed memento.
Each link followed uses the embedded datetime
  as the new target datetime (the selection from step 2 is
  forgotten), which introduces drift.
Additionally, it is unlikely that the selected URI-R was archived at the new
  target datetime; therefore, one or more additional redirects, each introducing
  drift, will be required before the best memento is displayed.

\end{enumerate}

Browsing using the \textit{Sliding Target} policy introduces
  two kinds of drift:
  \textit{Memento drift} by redirection to the actual memento and
  \textit{Target drift} by changing the target datetime.

\subsection{Sticky Target (Memento API)}
\label{ss:BrowsingViaMemento}

Browsing the Internet Archive using the Memento API uses the {Sticky Target}
policy.
The Sticky Target policy also introduces drift; however, the drift is
  constrained because the target datetime is maintained.
Here is the browsing process using Firefox and the MementoFox add-on:

\begin{enumerate}

\item \textbf{Select URI-R}.
Open Firefox and enable MementoFox.
Move the requested datetime slider to the desired target datetime.
All URI-Rs entered in the address bar or followed through clicking a link, are
  now dereferenced using the Memento API and redirected to the \textit{best}
  URI-M, which is the URI-M with Memento-Datetime closest to the target
  datetime.
Figure \ref{f:mementofox} shows the ODU Computer Science home for
  2005-05-15T00:28:03Z as dereferenced by MementoFox.
Drift is introduced when the target datetime redirects to a memento with a
  Memento-Datetime that is not the target datetime.

\begin{figure}[h!]
  \centering
  \includegraphics[width=3.25in]{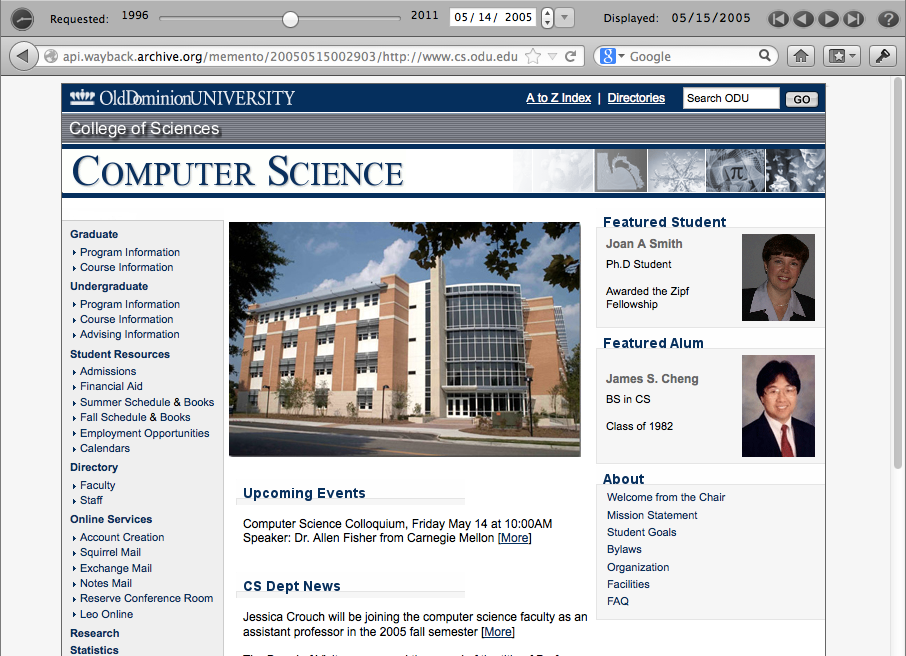}
  \caption{Memento API and MementoFox}
  \label{f:mementofox}
\end{figure}

\item \textbf{Browse additional URI-Rs}.
Each subsequent link clicked uses the target datetime selected by the slider.
Drift continues to be introduced by redirects as in step 1; however, using
  MementoFox and the Memento API allows the target datetime to remain the same
  for every request.

\end{enumerate}

Thus, browsing using the Memento API introduces only memento drift and does not
  introduce target drift.

\section{Experiment}
\label{s:Experiment}

\subsection{Samples}
\label{ss:Samples}

Building on our previous coverage work, we use the same four URI sample sets (DMOZ,
  Delicious, Bitly, and Search Engines) as in \cite{ainsworth-jcdl11}.
Each sample contains 1,000 randomly-selected URIs for 4,000 URI total.
URI random selection details can be found in \cite{ainsworth-arXiv:1212.6177}.

\begin{table}[h]
\centering
\caption{Archival Rates}
\label{t:ArchivalRates}
\begin{tabular}{lrrr}
  \hline
  Sample & \multicolumn{1}{c}{2013}
         & \multicolumn{1}{c}{2011} \\
  \hline
  DMOZ          & 95.2\% & 79\% \\
  Delicious     & 91.9\% & 68\% \\
  Bitly         & 23.5\% & 16\% \\
  Search Engine & 26.4\% & 19\% \\
  \hline
  Aggregate     & 59.4\% & 46\% \\
  \hline
\end{tabular}
\end{table}

Table \ref{t:ArchivalRates} shows the percentage of URI-Rs in
  the sample that were found to be archived during the experiment.
There are several notable differences from our 2011 results
  \cite{weigle-wsdlblog-110623}.
The DMOZ and Delicious samples are archived at a considerably higher rate;
  the Bitly and Search Engine samples rate are only slightly higher.
We attribute this to the increased archive availability provided
  by the Internet Archive over the past
  two years \cite{kahle-iablog-130109}.

\subsection{Procedure}
\label{ss:Procedure}

The examination of temporal drift was accomplished in January 2013.
To ensure adequate number of successful walks, 200,000 walks were attempted.
Each walk was accomplished in three phases:

\squishlisttwo
  \item Obtain the initial memento,
  \item Follow links, and 
  \item Calculate drift and statistics.
\squishend

Each walk iterates through the process of selecting URI-Rs and downloading
  mementos until either 50 steps are successful or an error is encountered.
The details of each walk step are captured, including steps that encounter
   errors.
The last step will contain the stop reason, unless the walk completed 50
  successful steps (in which case there is no stop reason).
The vast majority of walks encounter an error before reaching step 50.
The length of a walk is the number of successful steps.
For example, a walk that stops on walk step 6 (i.e. step 6 encounters an
  error), is length 5 because the first 5 steps were successful.
Table \ref{t:Definitions} defines the symbols used in the procedure description.

To ensure repeatability, a set of 200,000 random numbers (one per walk) were
  generated.
These random numbers were used as both the walk identifier and as the seed to
  initialize the walk's random number generator.
The experiment was run under Apple OS X 10.8.2 using Python 2.7.2 (the
  version included with OS X).  The built-in Python random number
  generator, \texttt{random.Random}, was used.
Resources were downloaded using \textit{curl}, which is much more robust than
  the Python libraries.
 
\begin{table}[h]
\centering
\caption{Definitions}
\label{t:Definitions}
\begin{tabular}{p{0.35in}p{2.65in}}
  \hline
  Term  & Definition \\
  \hline
  $W$   & An acyclic walk. \\
  $w$   & An acyclic walk index. \\
  $i$   & A walk step index. \\
  $t$   & The target datetime. $t_i$ is the target for walk step $i$. \\
  $R$   & A URI-R. $R_i$ is the $R$ selected for walk step $i$. \\
  $M$   & A URI-M. \\
  $M^a$ & A Memento API URI-M. $M^a_i$ is the $M^a$ for walk step $i$. \\
  $M^w$ & A Wayback Machine UI URI-M. $M^w_i$ is the $M^w$ for walk step $i$. \\
  $L $   & The set of link URI-Rs in a memento (dereferenced URI-M). \\
  $\mathcal{T}(M)$
      & The Memento-Datetime of $M$. \\
  $\mathcal{L}(M)$
      & The set of link $R$s in the memento identified by $M$. \\
  $\Delta(M)$
      & The drift of $M$ relative to the corresponding $t$.
        $\Delta(M_i) = |t_i - \mathcal{T}(M_i)|$ \\
  $\delta^a$
      & Memento API memento drift.  $\delta^a_i = \Delta(M^a_i)$. \\
  $\delta^w$
      & Wayback Machine UI memento drift.  $\delta^w_i = \Delta(M^w_i)$. \\
  \hline
\end{tabular}
\end{table}

\subsubsection*{Phase I. First Walk Step}

Phase I selects a walk's first URI-R, downloads the corresponding timemap,
  and downloads the first API and UI mementos.
Phase I accomplishes the first step of a walk.

\begin{enumerate}
  \item Randomly select $R_1$ from the 4,000 sample URIs.

  \item Retrieve the timemap for $R_1$ from the Internet Archive
using the Memento API.

  \item Randomly select a URI-M, $M_1$, from the timemap.
  $M_1$ yields this step's target datetime, $t_1 = \mathcal{T}({M_1})$.

  \item Dereference $M^a_1$ using $t_1$ from the IA Memento API.
  HTTP redirects may occur during dereferencing, yielding $M^{a\prime}_1$ as
    the final dereferenced URI-M.
  Note that $M^{a\prime}_1 = M^a_1$ may not hold.
  It follows that $\mathcal{T}(M^{a\prime}_1) = t_1$ also may not hold.
  This is the \textit{Sticky Target} policy.

  \item Calculate the corresponding $M^w_1$ and dereference it.
  As in step 4, HTTP redirects may occur during dereferencing, yielding
    $M^{u\prime}_1$ as the final dereferenced URI-M.
  In addition, the Wayback Machine returns {\it soft} redirects.
  These responses have HTTP status 200 but contain embedded JavaScript
    redirects; these are detected and followed. Note that
    $M^{w\prime}_1 = M^w_1$ may not hold.
  It follows that $\mathcal{T}(M^{w\prime}_1) = t_1$ also may not hold.
  This is the \textit{Sliding Target} policy.
\end{enumerate}

\subsubsection*{Phase II. Additional Walk Steps}

Phase II accomplishes a walk's second and subsequent steps.
It selects a link common to the API and UI mementos from the previous
  walk step and downloads the corresponding timemap and mementos.
If there are no common links, the walk stops.
In the following, $i$ is the current walk step.

\begin{enumerate}[resume]
  \item Extract the sets of link URI-Rs, $L^a = \mathcal{L}(M^a_{i-1})$ and
    $L^w = \mathcal{L}(M^w_{i-1})$, from the previous walk step's mementos.
  Denote the set of URI-Rs used in previous walk steps is $L^p$.
  Then, the set of common, usable URI-Rs for this walk step is
    $L^u_i = (L^a \cap L^w) - L^p$.
  Randomly select $R_i$ from $L^u_i$.

  \item Download the timemap for $R_i$ from the IA Memento API.

  \item Select the {\it best} URI-M, $M^a_i$, from the timemap.
  $M^a_i$ is the URI-M that minimizes $|t_1 - \mathcal{T}(M)|$, e.g. the URI-M
    nearest the initial target datetime.

  \item Dereference $M^a_i$ using $t_1$ from the IA's Memento API.
  As in step 4, HTTP redirects may occur during dereferencing, yielding
    $M^{a\prime}_i$.
  $M^{a\prime}_i = M^a_i$ and $\mathcal{T}(M^{a\prime}_i) = t_1$ may not hold.
  This is the \textit{Sticky Target} policy.

  \item Calculate $M^w_i$ using $t_i = \mathcal{T}(M^{u\prime}_{i-1})$, as the
    target datetime.
  Dereference $M^w_i$.
  As in step 5, HTTP redirects and {\it soft} redirects may occur,
    yielding $M^{w\prime}_1$ as final.
  Again, $M^{w\prime}_i = M^w_i$ and $\mathcal{T}(M^{w\prime}_i) = t_i$ may
    not hold.

  \item Repeat Phase II until $L^u_i$ is empty, an error occurs, or 50 walk steps
    have been completed successfully.
\end{enumerate}

\subsubsection*{Phase III. Drift Calculations}

Phase III calculates drift and statistics, ignoring duplicate walks.
Duplicate walks occurs for a number of reasons.
A common reason is failure on the first walk step because $R_1$ has never
  been archived.
A limited number of links or mementos is another reason.

\begin{enumerate}[resume]
  \item Calculate API drift, $\delta^a_i = \Delta(M^{a\prime}_i)$, for each
    successful walk step.
  Calculate the API drift mean, median, and standard deviation for the entire
    walk.

  \item Calculate Wayback Machine drift, $\delta^w_i = \Delta(M^{w\prime}_i)$,
    for each successful walk step.
  Calculate the Wayback Machine drift mean, median, and standard deviation for
    the entire walk.
\end{enumerate}

\subsection{Results}
\label{Results}

The 200,000 acyclic
  walks attempted resulted in 53,926 unique
  walks, of which 48,685 had at least 1 successful step.
Overall there were 240,439 successful steps, with an average of
  3.85 successful steps per walk.
Table \ref{t:WalksAndSteps} summarizes walks and steps by sample.

\begin{table}[h]
\centering
\caption{Walks and Steps}
\label{t:WalksAndSteps}
\begin{tabular}{@{} l @{ | } r @{ } r @{ } r @{ } r @{ | } r @{}}
  \hline
  \multicolumn{1}{@{}c@{ | }}{}
    & \multicolumn{1}{@{}c}{DMOZ}
    & \multicolumn{1}{@{}c}{S.Eng.}
    & \multicolumn{1}{@{}c}{Delicious}
    & \multicolumn{1}{@{}c@{ | }}{Bitly}
    & \multicolumn{1}{@{}c@{}}{Total} \\
  \hline
  Steps           &  64,661 &  26,047 & 124,020 &  25,711 & 240,439 \\
  Succ. Steps     &  48,445 &  20,177 &  98,560 &  20,189 & 187,371 \\
  ~w/$\delta^u>1\textrm{yr}$ &   1,761 &   1,212 &   3,028 &     700 &   6,701 \\
  ~w/$\delta^u>5\textrm{yr}$ &      75 &      13 &      16 &       7 &     111 \\
  \hline
  Unique Walks    &  16,221 &   5,873 &  25,482 &   5,524 &  53,100 \\
  Succ. Walks     &  15,009 &   4,604 &  24,451 &   4,621 &  48,685 \\
  Pct. Succ.      &  92.5\% &  78.4\% &  96.0\% &  83.7\% &  91.7\% \\
  \hline
  Mean Succ.      & \multirow{2}{*}{3.2} & \multirow{2}{*}{4.4} &
    \multirow{2}{*}{4.0} & \multirow{2}{*}{4.4} & \multirow{2}{*}{3.8} \\
  Steps/Walk & & & & & \\
  \hline
\end{tabular}
\end{table}

\subsubsection{Walk Length and Stop Causes}

\begin{figure}[h!]
  \centering  
  \includegraphics[width=3.25in, trim=0.033in 0.15in 0.25in 0.5in, clip=true]{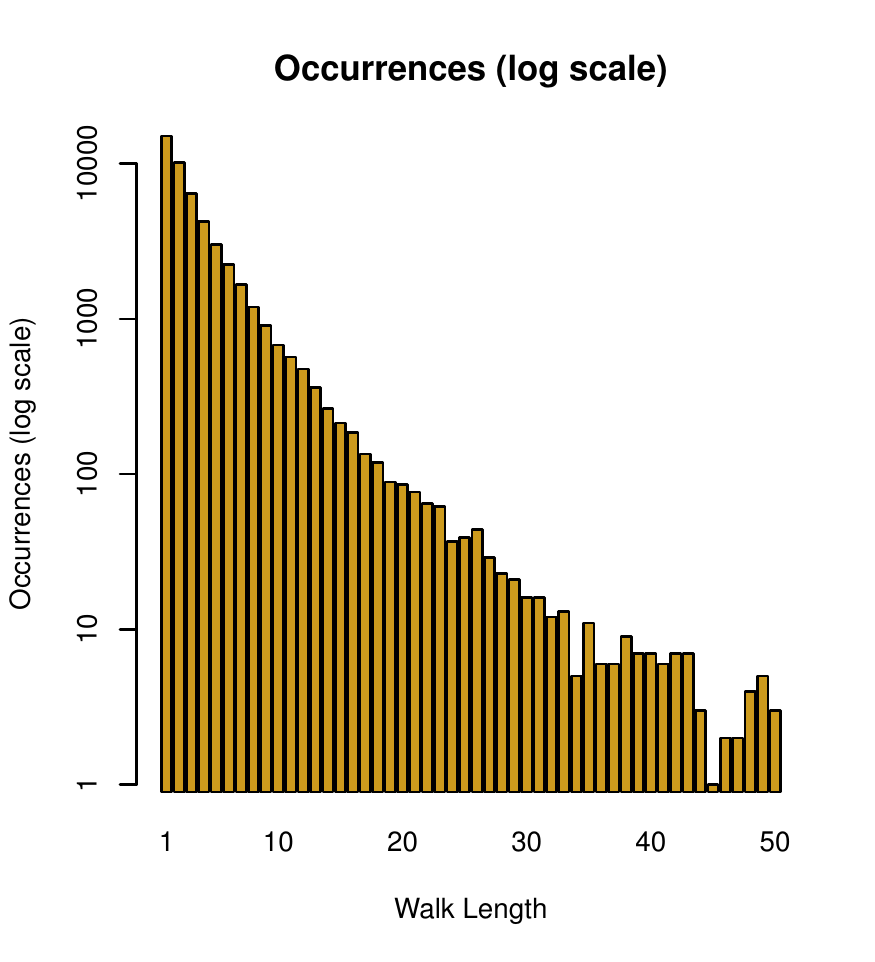}
  \caption{Occurrences by Walk Length}
  \label{f:occurrences:bylength:all}
\end{figure}

Figure \ref{f:occurrences:bylength:all} shows the distribution of
  successful walks by length.
(Note that \textit{Occurrences} is a log scale.)
Table \ref{t:occurrences:bylength} shows the details behind
  Figure \ref{f:occurrences:bylength:all}, broken out by sample.
Walks greater than 25 in length are summarized in groups of 5.
The number of steps decreases exponentially as walk length increases.
  Less than 1\% of walks progress past step 21.
For DMOZ, less than 1\% progress step past 19; Search Engine, step 23; Delicious,
  step 23; and Bitly, step 24.

\begin{table}[h]
\centering
\caption{Occurrences by Length}
\label{t:occurrences:bylength}
\begin{tabular}{c|rrrr|r}
  \hline
  \multicolumn{1}{c|}{Walk}
    & \multicolumn{1}{c}{}
    & \multicolumn{1}{c}{Search}
    & \multicolumn{1}{c}{}
    & \multicolumn{1}{c|}{}
    & \multicolumn{1}{c}{} \\
  \multicolumn{1}{c|}{Length}
    & \multicolumn{1}{c}{DMOZ}
    & \multicolumn{1}{c}{Engine}
    & \multicolumn{1}{c}{Delicious}
    & \multicolumn{1}{c|}{Bitly}
    & \multicolumn{1}{c}{Total} \\
  \hline
    1 & 5,355 & 1,239 & 7,193 & 1,289 & 15,076 \\
    2 & 3,571 & 924 & 4,857 & 817 & 10,169 \\
    3 & 1,891 & 598 & 3,311 & 623 & 6,423 \\
    4 & 1,212 & 381 & 2,228 & 415 & 4,236 \\
    5 & 791 & 315 & 1,588 & 314 & 3,008 \\
    6 & 583 & 232 & 1,168 & 259 & 2,242 \\
    7 & 417 & 178 & 877 & 186 & 1,658 \\
    8 & 258 & 153 & 651 & 136 & 1,198 \\
    9 & 187 & 111 & 498 & 108 & 904 \\
   10 & 144 & 79 & 377 & 79 & 679 \\
   11 & 114 & 71 & 306 & 74 & 565 \\
   12 & 99 & 51 & 279 & 48 & 477 \\
   13 & 72 & 44 & 200 & 46 & 362 \\
   14 & 54 & 32 & 144 & 35 & 265 \\
   15 & 39 & 30 & 119 & 26 & 214 \\
   16 & 33 & 26 & 108 & 20 & 187 \\
   17 & 20 & 23 & 76 & 18 & 137 \\
   18 & 24 & 14 & 68 & 12 & 118 \\
   19 & 19 & 12 & 46 & 12 & 89 \\
   20 & 14 & 10 & 47 & 15 & 86 \\
   21 & 20 & 11 & 36 & 9 & 76 \\
   22 & 7 & 13 & 28 & 16 & 64 \\
   23 & 11 & 11 & 33 & 7 & 62 \\
   24 & 7 & 4 & 22 & 4 & 37 \\
   25 & 8 & 3 & 25 & 3 & 39 \\
    26--30 & 27 & 18 & 68 & 20 & 133 \\
    31--35 & 7 & 7 & 30 & 14 & 58 \\
    36--40 & 6 & 3 & 23 & 3 & 35 \\
    41--45 & 6 & 2 & 14 & 2 & 24 \\
    46--50 & 6 & 3 & 6 & 1 & 16 \\
  \hline
  Totals & 15,002 & 4,598 & 24,426 & 4,611 & 48,637 \\
  \hline
\end{tabular}
\end{table}

Table \ref{t:StopCauses} summarizes the reasons walks stop before reaching
  step 50, split by timemap and memento.
Because the selection processes for the first and subsequent mementos
  differ, separate statistics are shown.
The stop causes are dominated by 403s, 404s, 503s, and \textit{No Common Links}.
The 403s are generally an archived 403; the original URI-R returned a 403
  when accessed for archiving.
The timemap 404s indicate that the URI-R is not archived.
Memento 404s can have two meanings: either the original URI-R returned
  a 404 when it was accessed for archiving or the memento has been redacted,
  i.e.\ removed from public access.
The 503s most likely indicate a transient condition such as an unavailable
  archive server, thus there is a chance that on repetition the resource will be
  found.
Resources that returned 503s were retried a single time
  one week after the first 503 was received.
Less than 1\% succeed on the second try.
\textit{Download failed} indicates that \textit{curl} was unable to download the
  resource; like the 503s, these were retried once.
\textit{Not HTML} indicates that the downloaded resource
  was not HTML and therefore not checked for links.
\textit{No common links} indicates that although both Memento API and Wayback
  Machine UI mementos were found, content
  divergence due to drift caused the two mementos to have
  no common links.

\begin{table}[h!]
\centering
\caption{Stop Causes}
\label{t:StopCauses}
\begin{tabular}{l|r@{\,}r|r@{\,}r}
  \hline
    & \multicolumn{2}{c|}{First Step}
    & \multicolumn{2}{c}{Other Steps} \\
  \cline{2-5}
  Stop Cause & Count & Percent & Count & Percent \\
  \hline
  \multicolumn{5}{l}{\textbf{Timemaps}} \\
  \hline
  HTTP 403 & 74 & 1.7\% & 4,803 & 9.1\% \\
  HTTP 404 & 1,327 & 30.1\% & 15,850 & 29.9\% \\
  HTTP 503 & 0 & 0.0\% & 43 & 0.1\% \\
  Other & 2 & 0.0\% & 180 & 0.3\% \\
  \hline
  \multicolumn{5}{l}{\textbf{Mementos}} \\
  \hline
  HTTP 403 & 52 & 1.2\% & 476 & 0.9\% \\
  HTTP 404 & 215 & 4.9\% & 3,633 & 6.8\% \\
  HTTP 503 & 1,957 & 44.4\% & 10,535 & 19.9\% \\
  Download failed & 154 & 3.5\% & 589 & 1.1\% \\
  Not HTML & 514 & 11.7\% & 2,856 & 5.4\% \\
  No Common Links &  & 0.0\% & 12,957 & 24.4\% \\
  Other & 117 & 2.7\% & 1,128 & 2.1\% \\
  \hline
  \multicolumn{1}{l|}{\textbf{Totals}}
     & \textbf{4,412} &  & \textbf{53,050} &  \\
  \hline
\end{tabular}
\end{table}

\subsubsection{Drift}

\begin{figure}[t]
  \centering  
  \includegraphics[width=3.25in, trim=0.033in 0.15in 0.25in 0.5in, clip=true]{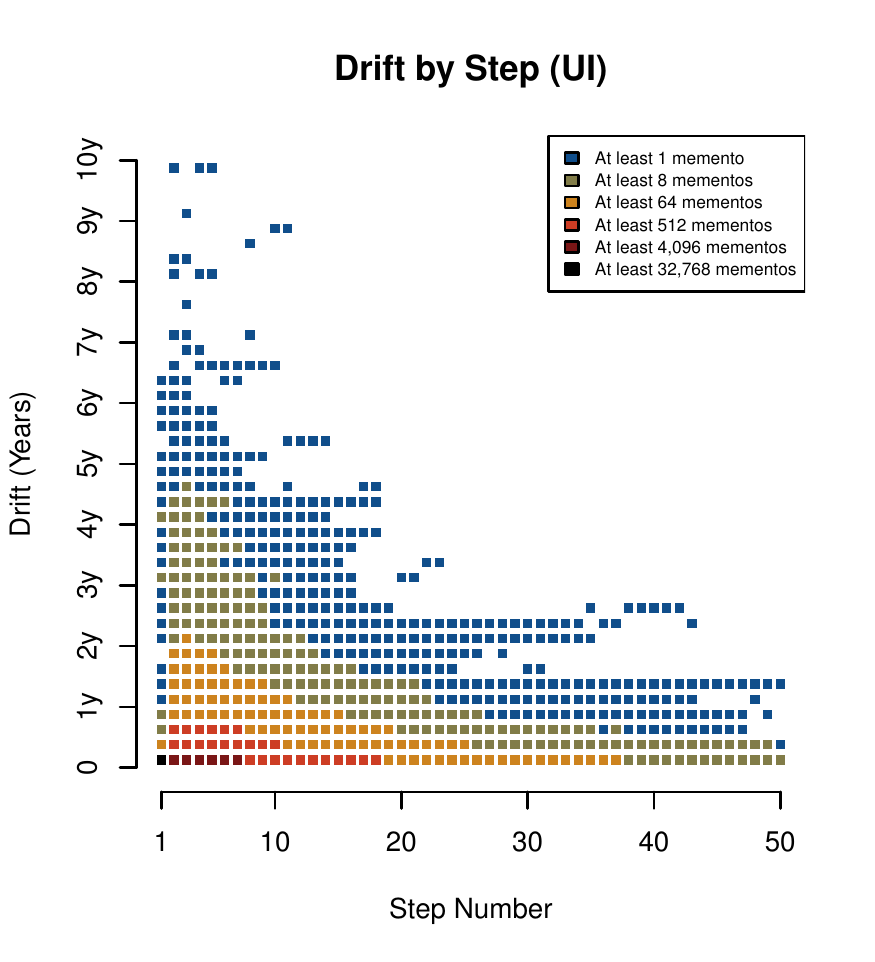}
  \caption{UI Drift by Step}
  \label{f:drift:all:ui}
\end{figure}

Figure \ref{f:drift:all:ui} illustrates the distribution of Wayback Machine
  mementos by drift ($\delta^w$).
The horizontal axis is the walk step number.
The vertical axis is the drift in years from the walk step's target datetime.
Color indicates memento density on an exponential axis.
As expected, density is highest for early steps and tapers off with
  as walk length increases.
Density is also highest at low drift values and many
  mementos appear to have very high drift.  However, only
  11,093 (4.6\%) exceed 1 year and only 172 (0.07\%) exceed 5 years
  (Table \ref{t:WalksAndSteps}).
It is also interesting that the first step shows drift
  (as high as 6.5 years).
The first target datetime is selected from a timemap,
  which sets the expectation that a memento for the datetime exists.
However, redirects (\ref{ss:Procedure} steps 4, 5, 9, and 11),
  from the URI-M in the timemap to the final URI-M cause
  drift---even on the first walk step.

\begin{figure}[t]
  \centering  
  \includegraphics[width=3.25in, trim=0.033in 0.15in 0.25in 0.5in, clip=true]{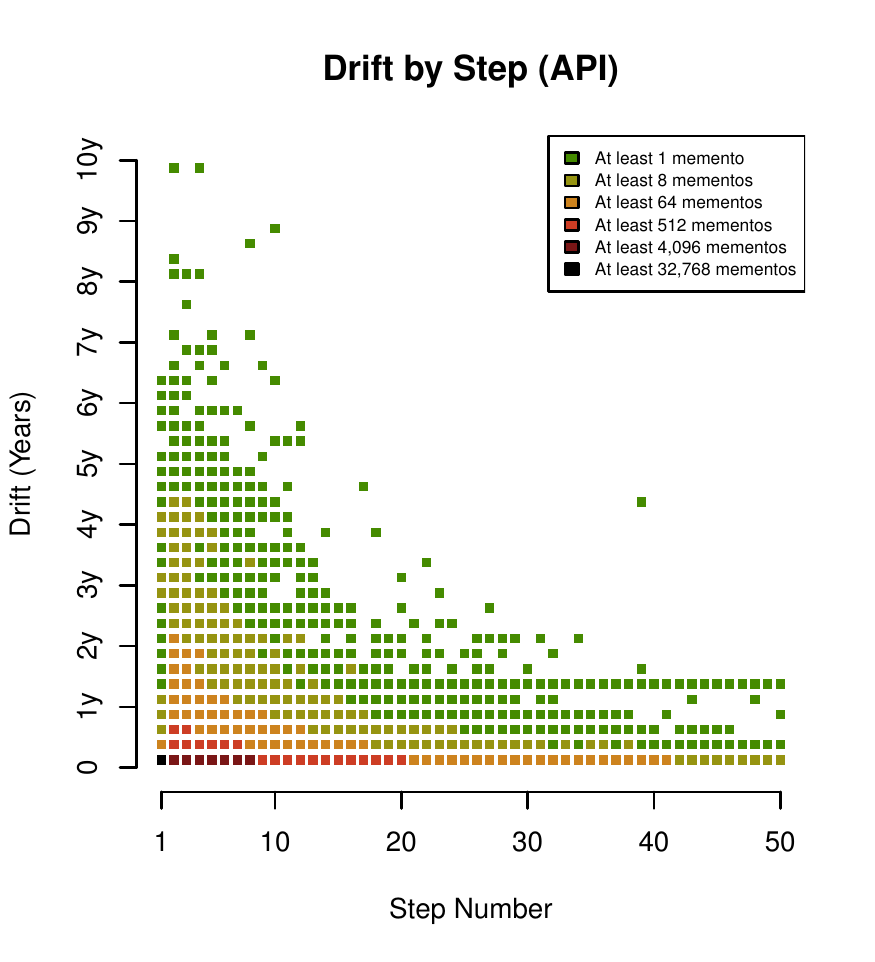}
  \caption{API Drift by Step}
  \label{f:drift:all:api}
\end{figure}

\begin{figure}[h]
  \centering  
  \includegraphics[width=3.25in, trim=0.033in 0.15in 0.2in 0.5in, clip=true]{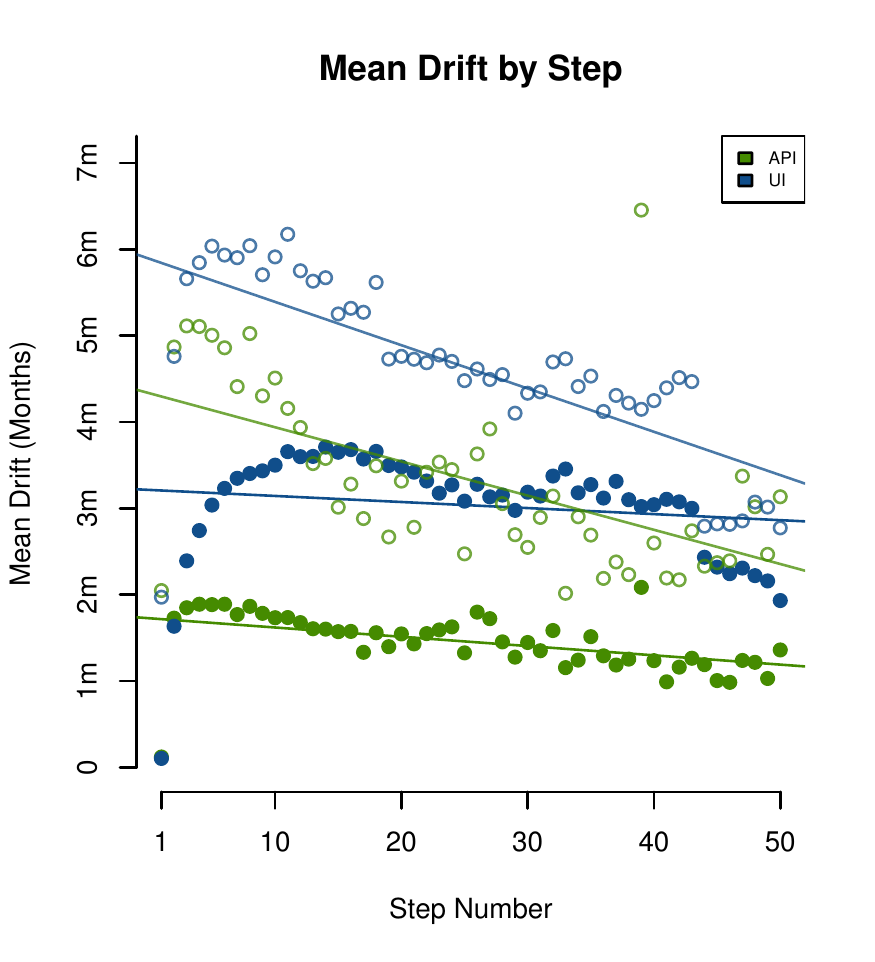}
  \caption{Standard Deviation Drift by Step}
  \label{f:drift:bystep:sd:all}
\end{figure}

Figure \ref{f:drift:all:api} illustrates the distribution of Memento API
  mementos by drift ($\delta^a$), which at first glance appears very
  similar to Figure \ref{f:drift:all:ui}.
Closer examination reveals that many mementos have lower drift
  when using the Memento API.
Figure \ref{f:drift:bystep:sd:all} shows the mean drift by step (solid circles) and
  standard deviation (hollow circles) for both the Memento API
  (green) and Wayback Machine URI (blue).
The Memento API, which uses the Sticky Target policy, results
  in 40--50 days less mean drift than the Sliding Target policy.
This delta appears to decrease above step 40, but there are only 40
  walks (0.0082\%) that achieve this many steps (see Table
  \ref{t:occurrences:bylength}), so the decrease is not significant.

\begin{figure}[h]
  \centering
  \includegraphics[width=3.25in, trim=0.033in 0.15in 0.25in 0.5in, clip=true]{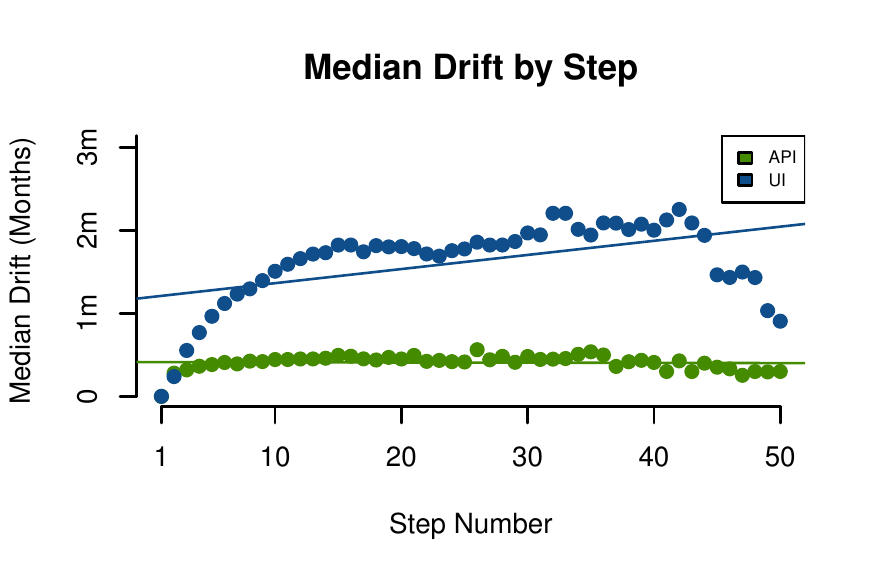}
  \caption{Median Drift by Step}
  \label{f:drift:bystep:median:all}
\end{figure}

Even below step 40, mean as
  a useful indicator of central tendency is in doubt.
As Figure \ref{f:drift:bystep:sd:all} shows, the
  standard deviation significantly exceeds the mean, particularly
  at low step numbers.
This indicates that median may be a better indicator of the central tendency.
Note the horizontal line of squares at 1.25 years
  in Figures \ref{f:drift:all:ui} and \ref{f:drift:all:api}.
Investigation revealed that well-archived, self-contained sites
  contribute more long walks than groups of loosely-linked sites.
For example, left column of every \url{http://www.101celebrities.com} page
  includes nearly 1,000 links to pages within the site; it is a primary
  contributor to the horizontal line.
The number of domains in a walk and their impact on the walk are
  discussed in \ref{ss:NumberOfDomains}.
The median is shown in Figure \ref{f:drift:bystep:median:all}.
The median shows lower average drift than the mean
  because it is less impacted by the outliers.
For this data, we believe median is the better measure of
  central tendency and will use median from here forward.

\subsubsection{Choice}
\label{ss:Choice}

\begin{figure}[h!]
  \centering  
  \includegraphics[width=3.25in, trim=0.033in 0.15in 0.25in 0.5in, clip=true]{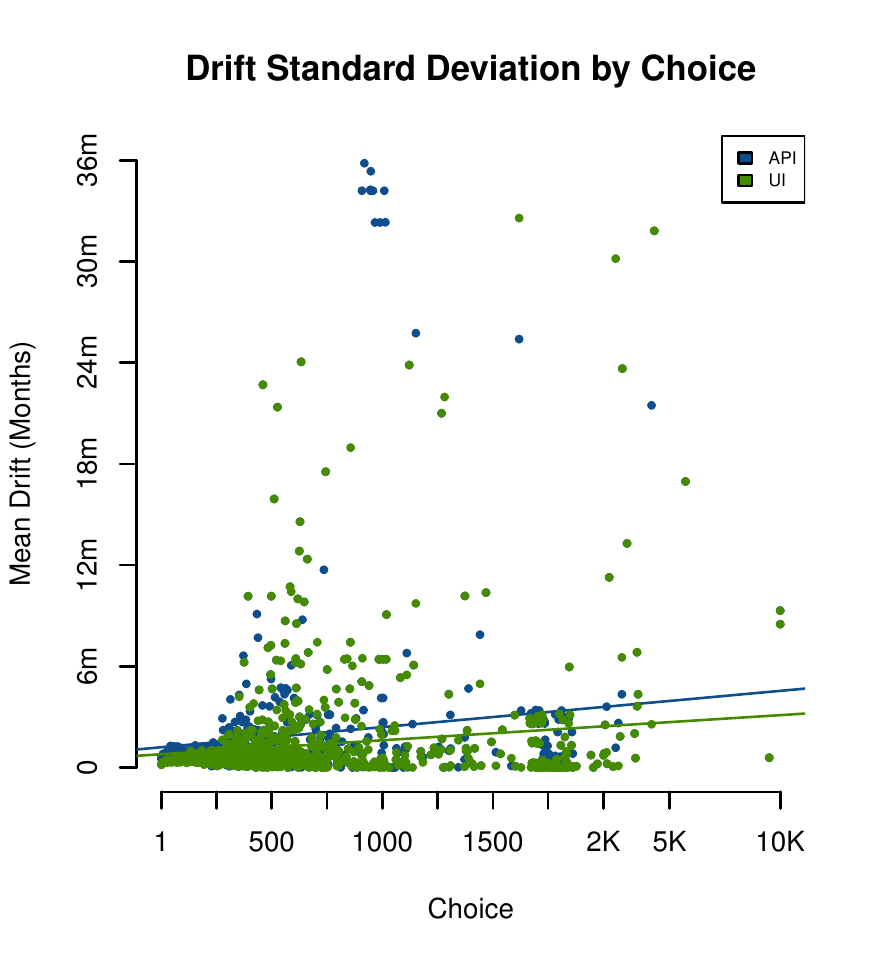}
  \caption{Median Drift by Choice}
  \label{f:drift:bychoice:median:all}
\end{figure}

Every walk step has a limited number of links to choose from.
Given a starting URI-R and Memento-Datetime, it is possible to represent all
  possible walks as a tree (\ref{ss:Procedure} step 6 disallows
  revisits).
Choice is the number of children common to both the Memento API and Wayback
  Machine mementos.
Figure \ref{f:drift:bychoice:median:all} shows the median drift by choice for
  the Memento API and Wayback Machine UI.
Clearly as choice increases, drift also increases.

\subsubsection{Number of Domains}
\label{ss:NumberOfDomains}

\begin{figure}[h!]
  \centering  
  \includegraphics[width=3.25in, trim=0.033in 0.15in 0.25in 0.5in, clip=true]{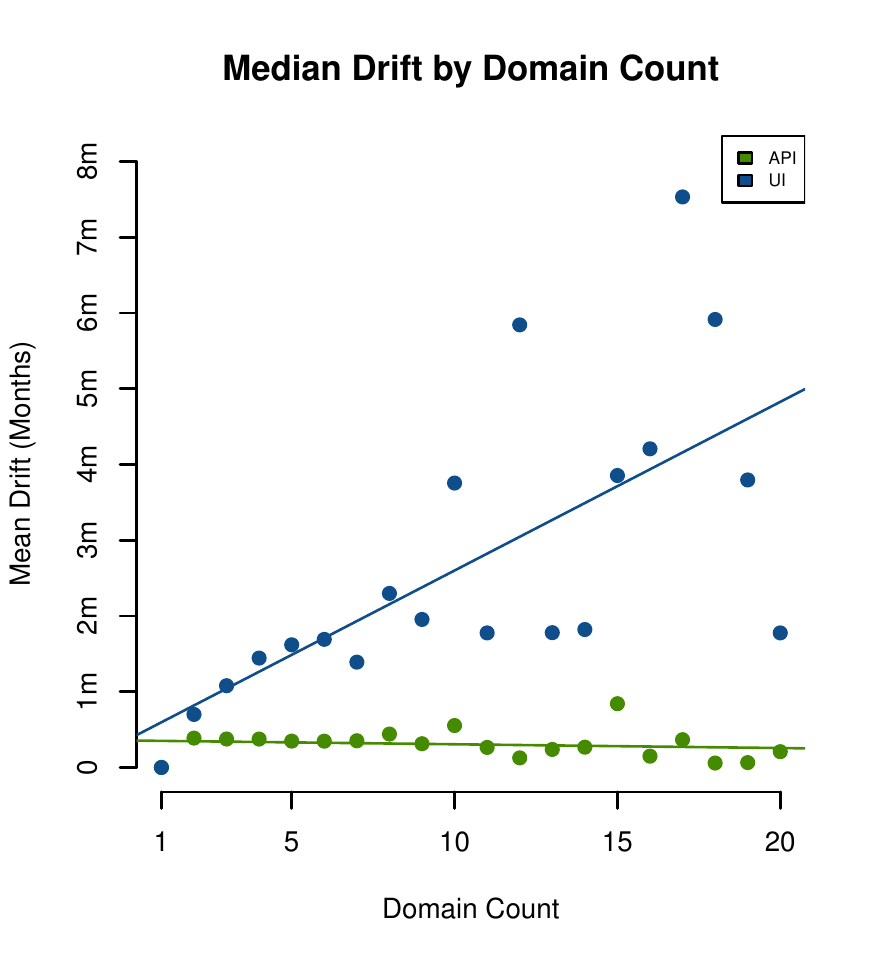}
  \caption{Median Drift by Number of Domains}
  \label{f:drift:bydomains:median:all}
\end{figure}

Through casual observations, we began to suspect that the number of
  domains accessed in a walk also impacted drift.
Figure \ref{f:drift:bydomains:median:all} graphs the relationship between
  the number of domains and drift.
The number of domains has a significant correlation with drift,
  but only for the Sliding Policy.

\subsubsection{Sample Differences}

In our 2011 research \cite{ainsworth-jcdl11}, we found that archival rates
  varied from 16\%--79\% (see Table \ref{t:ArchivalRates}) depending on the
  sample from which the URI-R originated.
This led to exploration of possible differences between acyclic walks
  based on sample.
We found there is not a significant drift difference based on sample source.

\subsubsection{Relaxed Shared URI Requirement}

\begin{figure}[h]
  \centering  
  \includegraphics[width=3.25in, trim=0.033in 0.15in 0.25in 0.5in, clip=true]{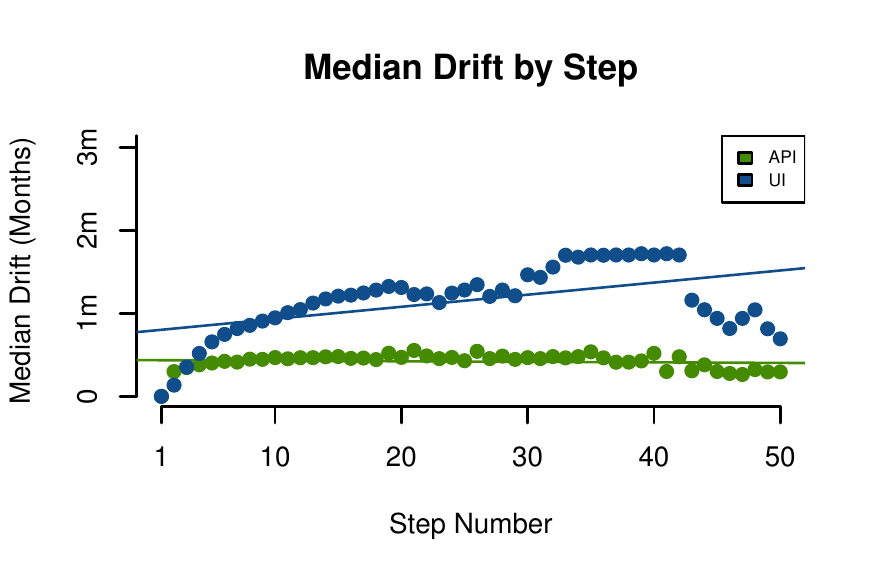}
  \caption{Median Drift by Step (Relaxed URI Requirement)}
  \label{f:drift:bystep:median:all:relaxed}
\end{figure}

An average walk length of 3.2 steps seems short.
Anecdotally, the authors' experience has been much longer walks when
  browsing the Internet Archive.
Much of this difference is likely due to the random rather than human
  browsing approach \cite{alnoamany-jcdl13}, but questions arose about
  requiring common URI-Rs at every walk step (\ref{ss:Procedure} step 6).
The experiment was run again using the same
  sample URI-Rs and random numbers while relaxing the requirement.
When a common URI-R was not available, two different URI-Rs were selected.
The results are summarized in Table \ref{t:WalksAndStepsRelaxedSummary}.
Compared with Table \ref{t:WalksAndSteps}, there is little change.
The number of steps and successful steps increased about 5\% each.
The number of unique and successful walks only increased by
  about 2.5\% and the average number of successful steps per walk increased
  by only 2.3\%.
Figure \ref{f:drift:bystep:median:all:relaxed} shows the median drift by step
  after relaxing the shared URI requirement; it is very similar to
  Figure \ref{f:drift:bystep:median:all}.
API drift is essentially the same and UI drift slightly reduced.
Even though relaxing the shared URI requirement reduces comparability between
  the two policies, the results also
  show that the Sticky Target policy
  controls temporal drift and that drift grows under the the Sliding Target policy.

\begin{table}[h]
\centering
\caption{Change in Walks and Steps w/Relaxed Shared URI Requirement}
\label{t:WalksAndStepsRelaxedSummary}
\begin{tabular}{l|rr|r}
  \hline
  \multicolumn{1}{c|}{}
    & \multicolumn{1}{c}{Strict}
    & \multicolumn{1}{c|}{Relaxed}
    & \multicolumn{1}{c}{Change} \\
  \hline
  Steps           & 240,439  & 251,439 &   +4.6\% \\
  Succ. Steps     & 187,371  & 196,999 &   +5.1\% \\
  ~w/$\delta^u>1\textrm{yr}$ &   6,701  &   6,344 &   -5.3\% \\
  ~w/$\delta^u>5\textrm{yr}$ &     111  &     118 &   +6.3\% \\
  \hline
  Unique Walks    &  53,100  &  54,474 &   +2.6\% \\
  Succ. Walks     &  48,685  &  50,043 &   +2.8\% \\
  Pct Succ.       &  91.7\%  &  91.9\% &   +0.2\% \\
  \hline
  Successful      & \multirow{2}{*}{3.2}  & \multirow{2}{*}{3.9} 
    & \multirow{2}{*}{+2.3\%} \\
  Steps/Walk & & & \\
  \hline
\end{tabular}
\end{table}

\section{Future Work}
\label{s:FutureWork}
We see several avenues of future work for this line of research.
The experiments conducted so far have focused on randomly-generated
  walks through a single archive.
AlNoamany et al.\ \cite{alnoamany-jcdl13} have looked at real-world walk
  patterns through analysis of the Internet Archive's web server logs.
Using these patterns to guide walks will provide more realistic
  link streams and result in temporal drift data more in line with actual
  user experience.
There are also domains that users tend to avoid, such as link farms, SEO,
  and spam sites.
Detecting and avoiding them, as a user would, will also move the data
  toward real-world user experience.
We also suspect that long walk drift is heavily influenced by clusters of
  closely-related domains and domains that primarily self-reference.
Applying an appropriate measure of clustering or similarity may shed some
  light on this topic.

Preliminary research has shown that the amount of drift can vary based on
  the starting date.
Repeating this study with a focus on the earliest and latest archived
  versions available will bear out (or disprove) our preliminary evidence.
Closely related to choosing the earliest or latest versions is starting with
  a variety of fixed datetimes.
In this case, we hypothesize increased first step drift for early dates
  followed by drift settling out after a few steps.

Recently, additional archives (e.g. the UK Web Archive) have
  implemented native Memento support and the Wayback Machine UI.
It will be interesting to see if other archives have temporal drift similar
  to the Internet Archive's.
Finally, the Memento architecture provides for aggregators, which are servers
  that combine the timemaps from multiple archives into a single, unified
  timemap.
The aggregators will make it possible to study drift across multiple archives.

\section{Conclusion}
\label{s:Conclusion}

We studied the temporal drift that occurs when browsing web archives
  under two different policies: Sliding Target and Sticky Target.
Acyclic walks through the Internet archived were conducted using the
  Memento API, which uses the Sticky Target policy, and the
  Wayback Machine UI, which employs the Sliding policy.
Measurements of drift were taken on three axis:
  number of steps (Table \ref{f:drift:bystep:median:all}),
  choice (Table \ref{f:drift:bychoice:median:all}), and
  number of domains (Table \ref{f:drift:bydomains:median:all}).
All three showed a showed a positive correlation with increased temporal
  drift for the Sliding Target policy.
For the Sticky Target policy, drift by step and drift by domain count
  showed no correlation.
Drift by choice showed low correlation for both policies; however, median
  drift for the Sticky Target was still lower overall.
The Sticky Target policy clearly achieves lower temporal drift.
Based on walk length, the Sticky Target policy generally produces at least
  30 days less drift than the Sliding Target policy.


\section{Acknowledgments}

This work supported in part by the NSF (IIS 1009392) and the Library of
  Congress.
We are grateful to the Internet Archive for their continued support of
  Memento access to their archive.
Memento is a joint project between the Los Alamos National Laboratory
  Research Library and Old Dominion University.

\bibliographystyle{abbrv}
\bibliography{references}
%
%

%
\balancecolumns
\end{document}